%% file: main.tex
\documentclass[a4paper,11pt]{article}
\usepackage{pos}
\usepackage{physics}
\usepackage{isotope}
\usepackage[utf8]{inputenc}

\input{macros}

\title{Preliminary study of the $H$ dibaryon in $\Nf=2+1$ lattice QCD}
\author[a]{André Bai\~ao Raposo}
\author[a]{John Bulava}
\author[b]{Jeremy R. Green}
\author[c]{Andrew D. Hanlon}
\author*[a]{Davide Laudicina}
\author[d,e]{Malcolm Lazarow}
\author[f]{Colin Morningstar}
\author[g]{Amy Nicholson}
\author[h]{Fernando~Romero-López}
\author[h]{Miguel Salg}
\author[d,e]{André~Walker-Loud}
\author[i]{Hartmut~Wittig}

\onbehalf{\\ For the Baryon Scattering (BaSc) collaboration}

\affiliation[a]{Institut f\"ur Theoretische Physik II, Ruhr-Universit\"at Bochum,
D-44780 Bochum, Germany}

\affiliation[b]{John von Neumann-Institut f\"ur Computing NIC, Deutsches Elektronen-Synchrotron DESY,\\
Platanenallee 6, 15738 Zeuthen, Germany}

\affiliation[c]{Department of Physics, Kent State University,
Kent, OH 44242, USA}

\affiliation[d]{Department of Physics, University of California,
Berkeley, CA 94720, U.S.A}

\affiliation[e]{Nuclear Science Division, Lawrence Berkeley National Laboratory
Berkeley, CA 94720, USA}

\affiliation[f]{Department of Physics, Carnegie Mellon University,
Pittsburgh, Pennsylvania 15213, USA}

\affiliation[g]{Department of Physics and Astronomy, University of North Carolina,
Chapel Hill, NC 27516-3255, USA}

\affiliation[h]{Institute for Theoretical Physics, Albert Einstein Center for Fundamental Physics,\\
University of Bern, 3012 Bern, Switzerland}

\affiliation[i]{Institute for Nuclear Physics, Johannes Gutenberg University Mainz,
55099 Mainz, Germany}

\emailAdd{andre.baiaoraposo@ruhr-uni-bochum.de}
\emailAdd{john.bulava@ruhr-uni-bochum.de}
\emailAdd{jeremy.green@desy.de}
\emailAdd{ahanlon7@kent.edu}
\emailAdd{davide.laudicina@ruhr-uni-bochum.de}
\emailAdd{mlazarow@berkeley.edu}
\emailAdd{fernando.romero-lopez@unibe.ch}
\emailAdd{miguel.salg@unibe.ch}
\emailAdd{walkloud@lbl.gov}
\emailAdd{wittigh@uni-mainz.de}

\abstract{We present preliminary results on the $I=0$, $S=-2$ $H$ dibaryon in $\Nf=2+1$ QCD. The calculation is performed with heavier-than-physical quarks ($m_\pi \approx 280$ MeV) on a
single CLS ensemble. Correlation matrices are constructed using the distillation technique and the three relevant channels, $\Lambda\Lambda$, $N\Xi$, $\Sigma\Sigma$, are investigated to determine the interacting spectrum relevant for $S$-wave across multiple momentum frames. The scattering amplitude is determined by solving the corresponding two-body quantization condition. These preliminary results are part of the ongoing efforts to determine the properties of di-hyperons and to establish whether the $H$ dibaryon exists down to physical quark masses.}

\FullConference{The 42nd International Symposium on Lattice Field Theory (LATTICE2025)\\
2-8 November 2025\\
Tata Institute of Fundamental Research, Mumbai, India\\}

\renewcommand{\hookAfterAbstract}{%
  \par\vspace{10pt}
  \hfill DESY-26-028
  \par\vspace{20pt}
}

\begin{document}
\maketitle

\input{introduction}

\input{setup}
\input{methods}

\input{results}
\input{summary}

\acknowledgments

The work of ABR, JB, DL is supported by the European Research Council (ERC) consolidator grant StrangeScatt-101088506.
The work of FRL and MS was supported in part by the Swiss National Science Foundation (SNSF) through grant No.~200021-236432, and the Platform for Advanced Scientific Computing (PASC) project “ALPENGLUE”.
The work of ML and AWL are supported in part by the U.S. Department of Energy (DOE), Office of Science, Office of Nuclear Physics, under grant contract numbers DE-AC02-05CH11231 and the DOE Topical Collaboration “Nuclear Theory for New Physics”, award No. DE-SC0023663.
C.J.M.~acknowledges support from the
U.S.~National Science Foundation (NSF) under award PHY-2514831. 

ABR and DL thank the Albert Einstein Center at the University of Bern for its
hospitality during a visit that significantly advanced this
work. 

Part of the calculations were performed on UBELIX (\url{https://www.id.unibe.ch/hpc}), the HPC cluster at the University of Bern. Part of the computations used a grant from the Swiss National Supercomputing Centre (CSCS) under project ID lp53 on Alps.
The contractions were performed using resources provided by the Gauss Centre for Supercomputing e.V.\ (\url{www.gauss-centre.eu}) on JUWELS~\cite{juwels} at the Jülich Supercomputing Centre, and the analysis on the HPC cluster Elysium of the Ruhr-Universit\"at Bochum, subsidized by the DFG (INST 213/1055-1).

\bibliographystyle{JHEP}
\bibliography{biblio}

%\begin{thebibliography}{99}
%\bibitem{...}
%....
%\end{thebibliography}

\end{document}

%% file: macros
\newcommand{\Nf}{N_{\rm f}}

\let\oldbibliography\thebibliography
\renewcommand{\thebibliography}[1]{%
  \oldbibliography{#1}%
  \setlength{\itemsep}{1.8pt}%
  \setlength{\parskip}{0pt}%
}

%% file: introduction.tex
\section{Introduction}
The existence of the $I=0$, $S=-2$ $H$ dibaryon remains one of the open questions in nuclear physics, and 
despite decades of experimental searches and extensive theoretical investigations, its existence has yet 
to be conclusively established. It is a six-quark state with $J^P=0^+$ and flavor content $uuddss$. It was first theorized by Jaffe in 1977, with the MIT bag model, as a deeply bound state with 
mass $M_H\approx 2100 \, {\rm MeV}$ and binding energy $B_H \approx 80 \, {\rm MeV}$ relative to the $\Lambda\Lambda$ threshold~\cite{PhysRevLett.38.617}.

From an experimental point of view its binding energy is constrained, by the so-called Nagara event, to be 
$B_H\lesssim 7 \, {\rm MeV}$ through the weak decay of the hypernucleus $\isotope[6][\Lambda\Lambda]{He}$~\cite{PhysRevLett.87.212502}. Despite the extensive research, there is still no conclusive 
evidence of the existence of such a state and, depending on its nature, the main decay channels are 
expected to be $H\to \Lambda p \pi^-\, , \Sigma^{-}p$ or $H\to \Lambda \Lambda \, , \Xi^{-} p$. The E224 and the E522 experiments at KEK~\cite{PhysRevC.75.022201,KEK-PSE224:1998trj} observed an enhancement in the $\Lambda\Lambda$ production near threshold which
is compatible with the presence of a resonance. More recently, measurements obtained by the ALICE Collaboration on pp and p-Pb collisions do not rule out the possibility of a shallow bound state with a binding energy of a few MeV~\cite{ALICE:2019eol}.

The $H$ dibaryon is also an interesting system for lattice QCD.  It was the first dibaryon system in which bound states were identified in calculations with dynamical quarks~\cite{PhysRevLett.106.162001,Inoue:2010es}.
Since that time, most lattice QCD calculations of the $H$ dibaryon have been performed at the $SU(3)$-symmetric point with heavier-than-physical pion masses. Despite the plethora of lattice calculations, there is, in general, no agreement on the binding energy even at the same pion mass. 
In Refs.~\cite{INOUE201228,Inoue:2010es}, the HALQCD Collaboration computed the 
binding energy of the $H$ dibaryon with three degenerate quarks for several values of the pion mass in the range $m_\pi \approx 470-1170 \, {\rm MeV}$ finding a binding energy ranging from $26$ to $\sim 50 \, {\rm MeV}$. An analogous calculation from the NPLQCD Collaboration at $m_\pi\approx 800 \, {\rm MeV}$ found a binding energy which is about twice the value obtained by HALQCD at the same pion mass~\cite{PhysRevD.87.034506}. In Ref.~\cite{Green:2021qol}, the $H$ dibaryon 
was investigated at several lattice spacings on CLS ensembles at
the $SU(3)$-symmetric point with a pion mass of about $400 \, {\rm MeV}$. The binding energy in the continuum limit is $B_H = 4.56 \, {\rm MeV}$. However at $a\approx 0.1 \, {\rm fm}$ it is about seven times larger, indicating that discretization errors play a crucial role in the interpretation of lattice results. The study of discretization errors at the same pion mass by employing different lattice formulations, is currently in progress~\cite{universality}.
Recently, a strong lattice spacing dependence has also been observed by the HALQCD Collaboration at $m_\pi \approx 800 \, {\rm MeV}$~\cite{halqcd2024}.

\begin{figure}[ht]
    \centering
    %\rule{0.8\textwidth}{0.4\textwidth}
    \includegraphics[width=0.48\linewidth]{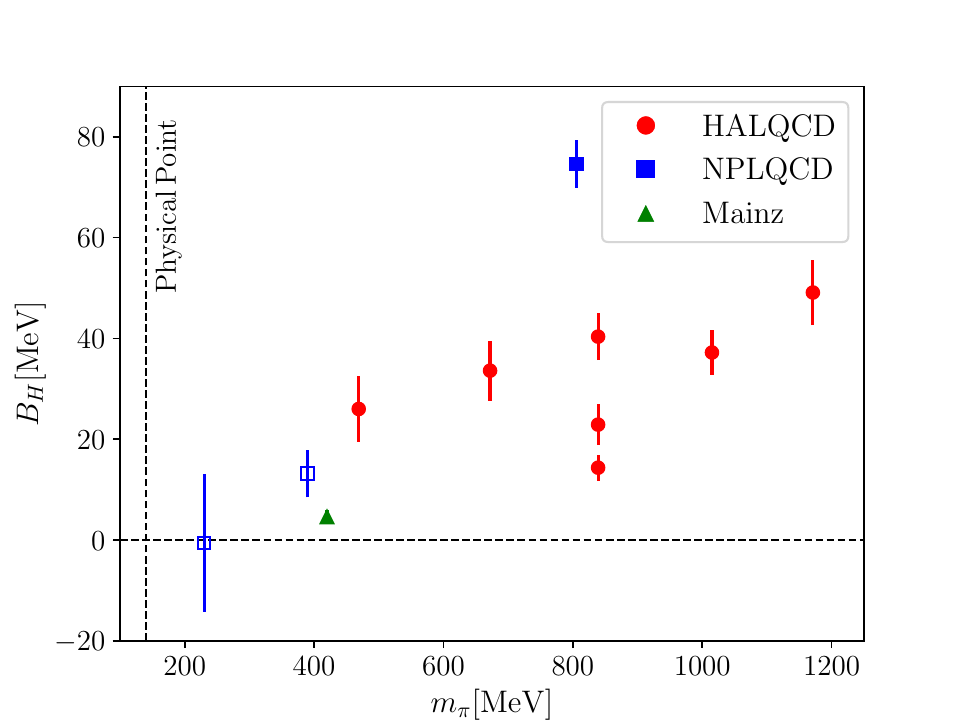}
    \caption{Lattice results for the binding energy as a function of the pion mass. Filled and empty symbols correspond to calculations in $SU(3)$-symmetric and $SU(3)$-broken QCD respectively.}
    \label{fig:history}
\end{figure}

Away from the $SU(3)$-symmetric point, employing an anisotropic setup, the NPLQCD Collaboration found $B_H \approx 13.2 \, {\rm MeV}$ at $m_\pi = 390 \,{\rm MeV}$ and a vanishing binding energy, but with large statistical uncertainty, at $m_\pi\approx 230 \, {\rm MeV}$~\cite{PhysRevLett.106.162001,Beane:2011zpa}. In Ref.~\cite{SASAKI2020121737} the HALQCD Collaboration performed simulations on a single lattice with $m_\pi$ close to its physical value including the $\Lambda\Lambda$ and the $N\Xi$ coupled channels finding only a weak interaction in the $\Lambda\Lambda$ sector. An overview of past lattice calculations performed by several collaborations at different pion masses and with different lattice setups is shown in Figure~\ref{fig:history}.

%% file: setup.tex
\section{Lattice setup}
This work is based on a single $ \Nf=2+1$ ensemble (D251) generated by the CLS initiative which lies on the $\Tr \left( m \right) = 2m_l + m_s = {\rm constant}$
trajectory. The gauge action has been discretized using the $O(a^2)$
improved L\"uscher-Weisz action, while the quark fields have been implemented using
$O(a)$-improved Wilson fermions with non-perturbatively tuned $c_{\rm sw}$.
The pion mass for this ensemble is heavier than physical, estimated 
to be $m_\pi \approx 280 \, {\rm MeV}$ and the large $m_\pi L \approx 6$ suppresses exponential corrections to the finite volume spectrum. 
For each of the 209 configurations we computed all-to-all smeared propagators employing sources on every single time slice. Further details on the ensemble used in this work are listed in Table~\ref{tab:D251}, while in Table~\ref{tab:single_baryon} we provide our determination of the pseudoscalar meson and octet baryon masses in this ensemble. The extraction of the masses has been carried out as detailed in Sec.~\ref{sec:spectrum}.
\begin{table}[ht]
    \centering
    \begin{tabular}{c c c c c c c c}
        \hline
        size & $a \, [{\rm fm}]$ & $m_\pi \, [{\rm MeV}]$ & $m_\pi L$ & $N_{\rm cnfg}$ & $N_{\rm tsrc}$ & $N_{\rm LapH}$\\
        \hline
         $64^3 \times 128 $& 0.064 & 280  & 5.91 & 209 & 128 & 160\\
        \hline
    \end{tabular}
    \caption{Details of the D251 CLS ensemble.}
    \label{tab:D251}
\end{table}

%% file: methods.tex
\section{Methodology}
\subsection{Operator construction}
\label{sec:ops}
We use a basis of bilocal baryon-baryon interpolating operators. As detailed in Refs. \cite{Green:2021qol,PadmanathMadanagopalan:2021exb} two-baryon operators are defined as products of momentum projected single baryon operators
\begin{align}
\label{eq:op}
\left[ B_1 B_2 \right]_{\vec{p}_1 , \vec{p}_2 } (t) = \sum_{\vec{x}_1,\vec{x}_2} e^{-i\vec{p}_1 \cdot \vec{x}_1} e^{-i\vec{p}_2 \cdot \vec{x}_2} \left[ qqq \right] (\vec{x}_1,t) \left[ qqq \right](\vec{x}_2,t) \, ,
\end{align}
where the single baryon operators have definite flavor structure. Starting from operators of the form in eq.~\eqref{eq:op}, for a given total momentum $\vec{P} = \vec{p}_1 + \vec{p}_2$ two-baryon operators are constructed to transform irreducibly under the representations of the little group of $\vec{P}$.
No local hexaquark operator has been included in the basis, since a previous study with $\Nf=2$ dynamical light quarks and quenched strange quark suggested that local hexaquark operators have small overlap with the flavor-singlet ground state and bilocal baryon operators are more effective in constraining the finite-volume spectrum ~\cite{PhysRevD.99.074505}.

Different momentum frames have been explored with $\vec{P}^2 = \left( 2\pi/L \right)^2 n$ and $n=0,1,2,3,4$ with single baryon momentum $\vec{p}_i^2 = \left( 2\pi/L \right)^2 n_i$ and $n_i = 0,\dots,5$. Correlation functions are constructed using distillation~\cite{PhysRevD.80.054506} and perambulators are computed with the \texttt{QUDA-lapH} code suite \cite{quda} which utilizes QUDA~\cite{Clark:2009wm} for multi-GPU, multi-grid solvers~\cite{Babich:2011np,Clark:2016rdz} and newly implemented routines for computing and projecting Laplacian-Heaviside (LapH) eigenvectors~\cite{Clark:2025cuz}. The total number of LapH eigenvectors used in this work is reported in the last column of Table~\ref{tab:D251}. Note that a balanced choice of the number of LapH eigenvectors is particularly critical. More smearing would lead to a stronger suppression of excited-state contamination at the cost of reduced statistical precision. Furthermore, $N_{\rm LapH}$ scales with the spatial volume and the computational cost of baryon contractions is proportional to $N_{\rm LapH}^4$, making the contraction step extremely expensive.
\subsection{Spectrum determination}
\label{sec:spectrum}
For each momentum frame and irreducible representation of the lattice symmetry group, correlation matrices $C_{ij}(t)$ are built using a large operator basis constructed as described in the previous section. For each $N\times N$ correlation matrix, we solve the corresponding generalized eigenvalue problem (GEVP)~\cite{LUSCHER1990222,Blossier:2009kd}. %This guarantees a stronger suppression of excited state contamination than the corresponding standard eigenvalue problem. 
We use the so-called single pivot method with reference and diagonalization times chosen such that the off-diagonal elements of the resulting rotated correlation matrix are compatible with zero within the statistical error for all time separations larger than the diagonalization time. The diagonal elements of the rotated correlation matrix provide approximations of the generalized eigenvalues, whose exponential behavior is dominated, for large separations, by the lightest $N$ states; see Ref.~\cite{Green:2021qol} for further details on state identification.

\begin{table}[h]
    \centering
    \begin{tabular}{c c c c c c}
    \hline
        $m_\pi \, [{\rm MeV}]$ & $m_K \, [{\rm MeV}]$ & $m_N \, [{\rm MeV}]$ & $m_\Lambda \, [{\rm MeV}]$ & $m_\Sigma \, [{\rm MeV}]$ & $m_\Xi \, [{\rm MeV}]$\\
        \hline
        $282.78(47)$ & $463.38(57)$ & $1044(2)$ & $1137(2)$ & $1182(2)$ & $ 1251(1)$\\
    \hline
    \end{tabular}
    \caption{Single hadron masses extracted from the D251 ensemble. The central values are obtained by model-averaging over different multi-exponential models and the error is obtained through bootstrap analysis; the scale setting error is omitted. 
    %See main text for further details on the model-averaging procedure.
    }
    \label{tab:single_baryon}
\end{table}

\begin{figure}[t]
    \centering
    %\rule{0.8\textwidth}{0.4\textwidth}
    \includegraphics[width=0.45\linewidth]{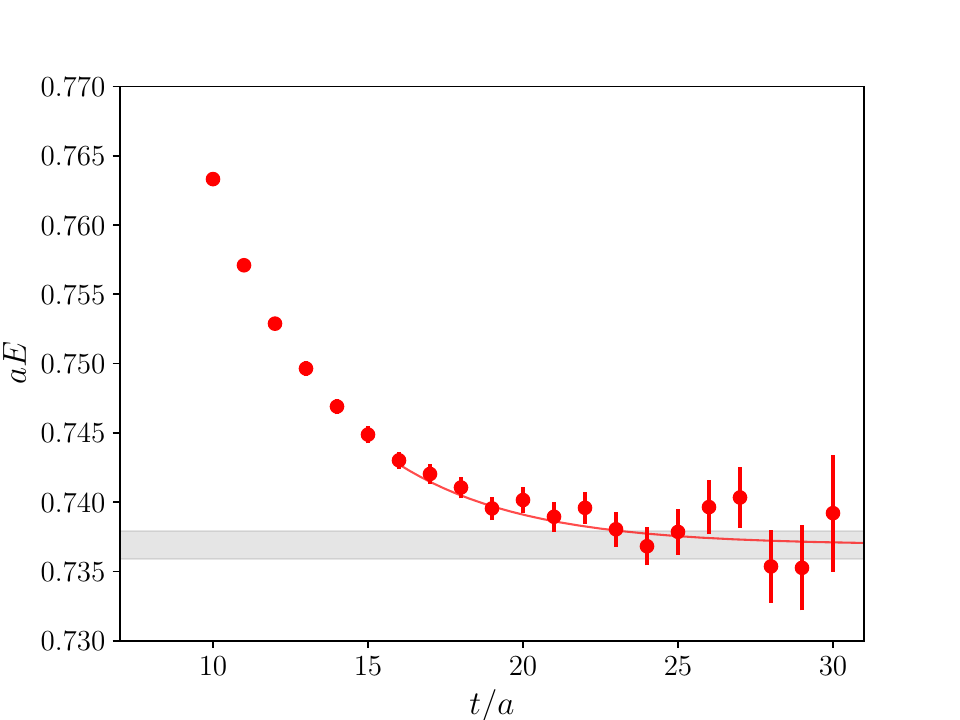}
    \includegraphics[width=0.45\linewidth]{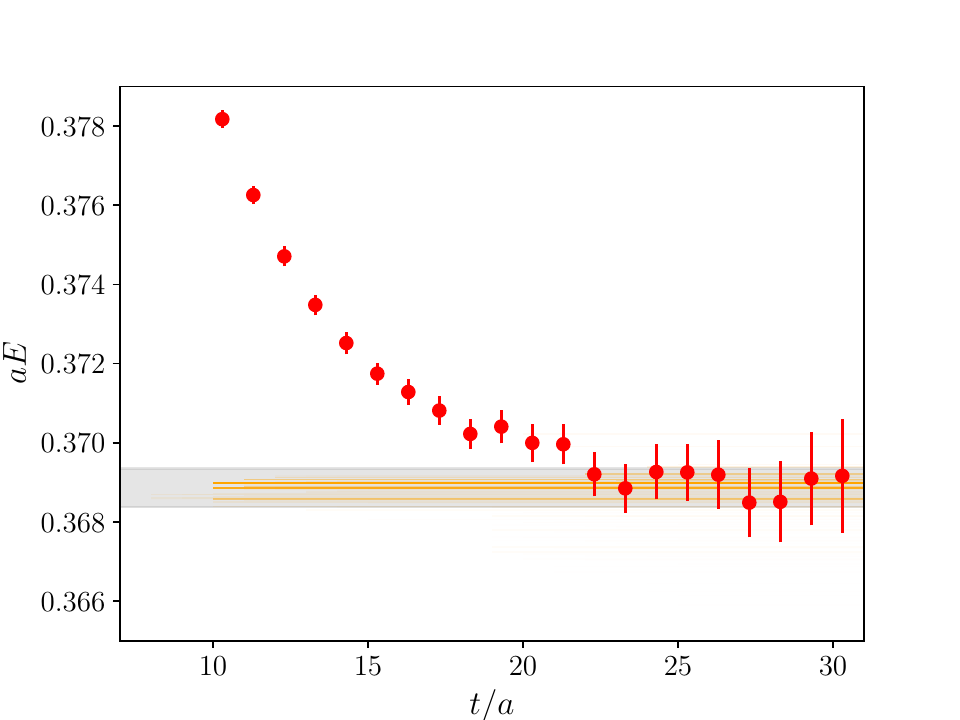}
    \caption{Left: effective energy of the ground state two-particle correlator in the rest frame for the $A_{1g}$ irrep. The red curve is the optimal model chosen as discussed in the main text, while the gray band shows the estimate of the corresponding energy. Right: effective energy of the one-particle correlator associated with the $\Lambda$ baryon in the rest frame. The orange lines are different estimates of the mass, the transparency of each line corresponds to the weight of each model entering the model-averaging procedure, while the gray band is the final estimate.
    }
    \label{fig:effm}
\end{figure}

In order to avoid the problem of false plateaux which might appear in the extraction of the energy levels when fitting ratios of correlation functions, we perform multi-exponential fits directly to the diagonal elements of the rotated correlation matrix. We use an information theory-based criterion to select the optimal description of the dataset. Each correlator is fit with $n$-exponential models with $n=1,\dots,6$, for different choices of prior and GEVP parameters and $t_{\rm min}$~\cite{BaSc:2025yhy}. The best estimate for each energy level is then obtained by selecting the model and the dataset with the lowest 
Akaike Information Criterion (AIC) \cite{Neil:2023pgt}. In Figure~\ref{fig:effm} on the left we show the effective energy for the ground state in the rest frame in the $A_{1g}$ irrep, with the band corresponding to the estimated plateau from the optimal model. This strategy selects two-exponential fit functions as the best model in most of the cases, except for a couple of cases where the rotated correlators are better described by a three-exponential model. While this spectrum extraction method avoids the issue of false plateaux, it is not able to capture correlations between single and two-baryon correlators as more sophisticated fitting strategies allow \cite{BaSc:2025yhy}. The extraction of the interacting finite-volume spectrum with simultaneous fits of single and two-baryon correlators is currently in progress.

A similar strategy is used for the extraction of the single hadron energy levels in all momentum frames, with the only notable difference that the final estimate of the ground state energy is obtained by using a model-averaging procedure. This approach has the advantage that the systematic error induced by excited state contamination and arbitrary choices, e.g. $t_{\rm min}$ and prior hyperparameters, can be reliably assessed; see Ref. \cite{PhysRevD.103.114502} for further details on the procedure. In Figure~\ref{fig:effm} on the right we show the effective energy of the $\Lambda$ baryon in the rest frame. In Table \ref{tab:single_baryon} we report our determinations of the pseudoscalar meson and of the single baryon octet masses from the ensemble under study. In the future, we plan to extend the model-averaging procedure to the two-baryon correlator simultaneous fits as well.

\subsection{L\"uscher analysis}
\label{sec:FV}
Infinite volume scattering information is related to the finite volume spectrum through the L\"uscher quantization condition \cite{Luscher:1986pf,RUMMUKAINEN1995397,Briceno:2014oea}
\begin{align}
    \det_{\ell,m} \left[ F(E,\vec{P},L)^{-1} + {\cal K}(E_{\rm cm}) \right] = 0 \, ,
\end{align}
where $E_{\rm cm} = \sqrt{E^2-\vec{P}^2}$ is the center-of-mass energy. Here ${\cal K}(E_{\rm cm})$ is the infinite-volume $K$-matrix, while $F(E,\vec{P},L)^{-1}$ encodes the finite volume kinematics and explicitly depends on the finite-volume energy levels.

In the $SU(3)$-symmetric limit, two-baryon states can be classified in terms of irreducible representations of the flavor group as \cite{Inoue:2010hs}
\begin{align}
\label{eq:flavor_irreps}
\mathbf{8} \otimes \mathbf{8} = \mathbf{10} \oplus \mathbf{\overline{10}} \oplus \mathbf{8_S} \oplus \mathbf{8_A} \oplus \mathbf{27} \oplus \mathbf{1}    \, ,   
\end{align}
with the $H$ dibaryon belonging to the singlet representation. Being a flavor-symmetric channel, spin-zero energy levels are associated with even partial waves. Therefore, the lowest scattering channel to be considered is $^1 S_0$. Odd partial waves, which are spin one, factor out in the quantization condition and higher partial wave effects start contributing from $^1 D_2$.

Away from the $SU(3)$ symmetric point, the three relevant coupled channels are $\Lambda\Lambda$, $\Sigma\Sigma$ and $N\Xi$. The presence of non-identical particles in the $N\Xi$ channel induces mixing between irreps associated with $S$ and $P$ waves. Equivalently, the $\Lambda\Lambda$ and the $\Sigma\Sigma$ states only receive contributions from the flavor-symmetric irreps $\mathbf{1}$, $\mathbf{27}$ and $\mathbf{8_S}$, while the $N\Xi$ can be written in the flavor basis in terms of both symmetric and anti-symmetric flavor irreps \cite{Inoue:2010hs}.
In moving frames, finite-volume effects couple the lowest $N\Xi$ scattering channel $^1 S_0$ to $^1 P_1$, while, at the same time, the scattering amplitude couples $^1 P_1$ to $^3 P_1$. As a consequence, constraining the infinite volume $K$-matrix away from the $SU(3)$ symmetric point is a much more complicated task due to the presence of three distinct coupled channels, with the consequent mixing of different partial waves.

In Ref. \cite{Green:2021qol}, it was pointed out that another possible source of systematic error in the finite-volume analysis is the presence of the $t/u$-channel left-hand cuts in the scattering amplitude due to long-range interactions arising from the exchange of pseudoscalar mesons. Close to the left-hand cut, the standard L\"uscher formalism breaks down and, in order to correctly incorporate effects coming from light pseudoscalar-meson exchange, extensions to the standard formalism have to be taken into account \cite{Meng:2023bmz,Raposo:2023oru}.
In the $SU(3)$ symmetric regime, with sufficiently heavy pion mass, the only left-hand cut lies well below the $\Lambda \Lambda$ elastic threshold, and its effects can possibly be controlled by excluding energy levels lying too close to the cut. Away from the $SU(3)$ symmetric point, the situation is qualitatively different. Due to the presence of three coupled channels, multiple left-hand cuts, both from diagonal and cross channels, are present in the kinematical region between the lowest and the highest elastic thresholds. Since the $\Lambda \Lambda$ and the $N\Xi$ thresholds are substantially lower than the $\Sigma\Sigma$ threshold, see Table \ref{tab:single_baryon}, 
we also note the presence of a two-pion exchange left-hand cut in the diagonal $\Sigma\Sigma$ channel for which no extension to the standard formalism is currently available; its possible effects deserve further investigation.

%% file: results.tex
\section{Preliminary results}

\begin{figure}[ht]
    \centering
    %\rule{0.8\textwidth}{0.4\textwidth}
    \includegraphics[width=0.95\linewidth]{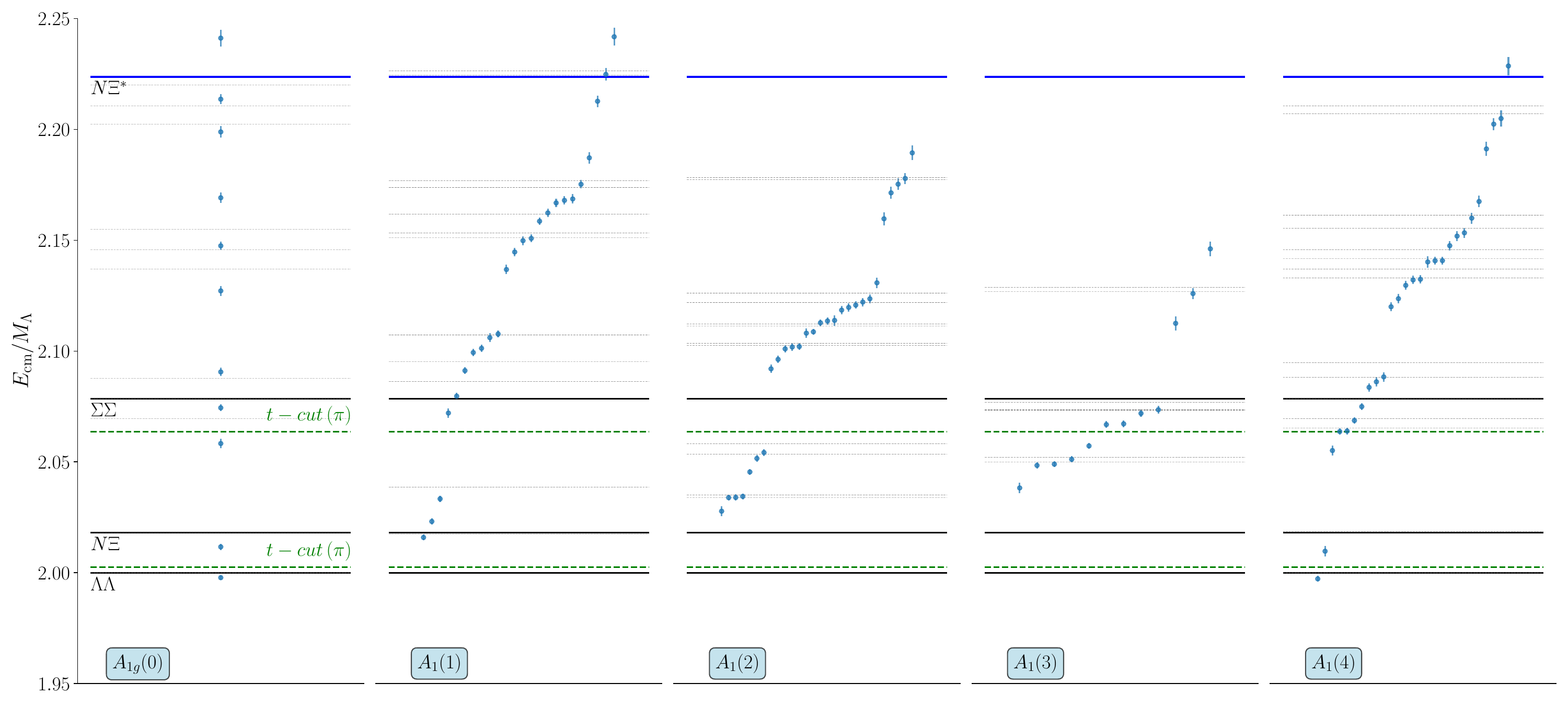}
    \caption{The spectrum on the D251 ensemble. Blue points are the energy levels extracted from lattice correlators, gray dashed lines are the corresponding non-interacting energy levels. Black and blue solid lines represent the relevant two-body thresholds. The green dashed lines are the most relevant left-hand cuts in the diagonal channels. Points are horizontally offset for clarity.
    }
    \label{fig:spectrum}
\end{figure}

In Figure \ref{fig:spectrum} we show the two-baryon energy spectrum in the center-of-momentum frame. Different panels correspond to different momentum frames and the energy levels displayed are the ones related to two-baryon interpolating operators which transform under the $A_{1g}(n)$ and $A_1(n)$ irreps of the cubic group, where $n=0,1,2,3,4$ is related to the magnitude of the total momentum $\vec{P}$ as discussed in Section~\ref{sec:ops}. All the energy levels are normalized by the mass of the $\Lambda$ baryon. Blue points correspond to the finite-volume energy levels extracted from the lattice correlators as detailed in Section~\ref{sec:spectrum}, gray dashed lines are the corresponding non-interacting levels and solid lines indicate the various two-particle thresholds. 

In the same plot we show the most relevant $t$-channel cuts due to one pion exchange for the diagonal channels. Note that the lowest left-hand cut is due to one $\eta$ meson exchange in the $\Lambda\Lambda\to\Lambda\Lambda$ channel and it lies well below the lightest energy level and, for this reason, it is not shown in the plot.
The two-pion exchange $t$-channel cut in the diagonal $\Sigma\Sigma$ channel is not shown in the plot. However, it is worth noting that its position lies just below the $N\Xi$ production threshold.

Given the discussion in Section ~\ref{sec:FV}, a complete L\"uscher analysis would involve three coupled channels, $S$ and $P$ wave mixing and a correct incorporation of (at least) one-pion exchange cuts. All these aspects will be taken into account in a future work. Here we present a preliminary analysis of the lowest-lying spectrum neglecting partial wave mixing and the presence of left-hand cuts. We restrict the finite-volume analysis to the six lowest energy levels in the rest frame. In the $A_{1g}$ irrep of the cubic group, the dominant contribution comes from the $^1S_0$ partial wave, which couples to flavor-symmetric states only. The next higher partial waves in this irrep, flavour-symmetric $^1 G_4$ and flavour-antisymmetric $^3 G_4$, are suppressed relative to $^1S_0$.

From the flavor decomposition in eq.~\eqref{eq:flavor_irreps}, the symmetric physical states $\Lambda\Lambda$, $\Sigma\Sigma$ and the symmetric component of the mixed state $N\Xi$, can be written as linear combinations of the flavor states ${\mathbf{1}},{\mathbf{8_s}}$ and ${\mathbf{27}}$ \cite{Inoue:2010hs}. Inspired by chiral effective theory, the interactions among different channels are split in terms of $SU(3)$ symmetric and $SU(3)$-breaking interactions. As a consequence, a simple parameterization for the inverse $K$-matrix is given by
\begin{align}
    {\cal K}^{-1} (E_{\rm cm}) = U \left( {\cal K}^{-1}_{SU(3)} + {\cal K}^{-1}_{\rm breaking} \right) U^{-1} \, ,
\end{align}
where the flavor-symmetric and the $SU(3)$-breaking terms read respectively
\begin{align}
\label{eq:K}
    {\cal K}^{-1}_{SU(3)} = \begin{pmatrix}
        c_{\mathbf{1}} & 0 & 0 \\
        0 & c_{\mathbf{8_s}} & 0 \\
        0 & 0 & c_{\mathbf{27}}
    \end{pmatrix} \, , \qquad
    {\cal K}^{-1}_{\rm breaking} = \begin{pmatrix}
        0 & c_{\mathbf{1\leftrightarrow 8_s}} & c_{\mathbf{1\leftrightarrow 27}} \\
        c_{\mathbf{1\leftrightarrow 8_s}} & 0 & c_{\mathbf{8_s\leftrightarrow 27}} \\
        c_{\mathbf{1\leftrightarrow 27}} & c_{\mathbf{8_s\leftrightarrow 27}} & 0
    \end{pmatrix} \, ,
\end{align}
while the unitary matrix $U$ contains $SU(3)$ Clebsch-Gordan coefficients which relate the flavor and the particle states. We employ a minimal five-parameter model where $c_{\mathbf{27}}$ and $c_{\mathbf{8_s}}$ are assumed to be constant, while $c_{\mathbf{1}}$ contains a linear dependence in the energy. The main reason for this choice is the fact that both lattice calculations and chiral effective theory arguments suggest the flavor singlet irrep to be responsible for attractive interaction~\cite{Haidenbauer:2011ah,Haidenbauer:2015zqb}. Similarly a direct inspection of the fitted $K$-matrix suggests that the $c_{\mathbf{1\leftrightarrow 27}}$ is the most relevant $SU(3)$-breaking term appearing in our parameterization. For this reason, in our model of the inverse $K$-matrix we explicitly set $c_{\mathbf{1\leftrightarrow 8_s}}$ and $c_{\mathbf{8_s\leftrightarrow 27}}$ to zero and leave $c_{\mathbf{1\leftrightarrow 27}}$ as a constant parameter to be fitted.

The parameterization in eq.~\eqref{eq:K}, despite being very simple, is able to describe the lowest-lying flavor-symmetric spectrum in the rest frame and the $SU(3)$-breaking term $c_{\mathbf{1\leftrightarrow 27}}$ turns out to be crucial to have a good quality-of-fit. It is however important to notice that when we include higher-lying energy levels in the analysis, the model in eq.~\eqref{eq:K} is not general enough to provide a good description of the lattice data. This is not surprising since this simple parameterization might not be able to fully capture effects arising from $SU(3)$ breaking. This is likely to be relevant given that $M_K^2 - M_\pi^2$ is large on this ensemble.
Furthermore, it is important to note that the quantization condition employed does not incorporate the effects of the left-hand cuts present in the studied kinematical region.

%% file: summary.tex
\section{Summary and outlook}
We presented our preliminary results on the $S=-2$, $I=0$ $H$ dibaryon on a single CLS ensemble away from the $SU(3)$ symmetric point. The interacting spectrum has been extracted over five different momentum frames from correlation matrices built from a large basis of operators constructed using the distillation technique. Our future plans include a more sophisticated determination of the spectrum using simultaneous fits to exploit correlations among single and two-baryon correlators and to use the model-averaging procedure on two-baryon correlators in order to obtain a reliable estimate of the systematic uncertainty. Furthermore, we are currently 
working on treating the effects of single-exchange left-hand cuts. As a final remark, let us stress that, given the large cut-off effects observed in previous lattice calculations, performing simulations at different lattice spacings will be crucial to correctly estimate the size of discretization effects.